# Non-conventional Superconductors and Percolation

Xuan Zhong Ni, and M. H. Jiang

*Santa Rosa, California, US*

(December 2010)

Abstract:   This paper presents a novel theory for understanding the mechanics behind non-conventional superconductors.   It presents the hypothesis that non-conventional superconductors are 2D lattices of super-cells and that the superconductivity mechanism involves the ordered hopping of itinerant electrons along with phase transitions in the edge percolations.

I. INTRODUCTION

This paper seeks to elucidate a novel theory in determining the mechanics behind non-conventional superconductors.   This theory involves three critical parts.

We first make the assertion that non-conventional superconductors are made up of 2-D lattices of super-cells.   Each super-cell is a large molecule composed of several physical cells of the lattice crystal itself.   For example, the large molecule in $CuO_2$ is an $O_4$ molecule (which is actually two physical cells of $CuO_2$).   In $Bi_2Se_3$, the large molecule is a hexagon in the xy-plane with two layers of Bi atoms and three layers of Se atoms for a total of nine $Se^{2-}$ ions and six $Bi^{3+}$ ions in one super-molecule.   For FeSe, it is possible that the basic molecule has a size of eight Fe-Fe distances (as observed in an STM image[1]).

Secondly, these molecules will have hybrid p-electrons.   Accordingly, the hybrid $p_z$ electrons form electron clouds on both sides of the physical ion layer.   These cloud electron wavefunctions could be either symmetric or anti-symmetric to the ion plane layer.   In $CuO_2$, the wavefunction is symmetric.   In $Bi_2Se_3$, it is anti-symmetric. And in FeSe, it is shifted symmetrically in the xy-direction.

Finally, these $\pi_{2pz}$ electrons shared by the large molecules in the crystal are itinerant electrons.   We theorize that superconductivity exists due to the ordered hopping of itinerant electrons to their nearest neighbors (n.n) along with phase transitions in the edge percolations.   The hopping occurs between the n.n in the cloud planes so that there are no collisions between the electrons and the ions.   The percolation, meanwhile, guarantees a physical infinite path when the pair connection between the n.n reaches a certain density.   This physical path will provide the superconducting path.

We have derived more new terms in the Hubbard model calculation by including a multiple center approximation.   The original Hubbard model includes only one-center integral in the electron-electron interaction due to the localized d-electron wavefunction.   In our case, the large molecular cells share their edge atoms with their n.n cells.   Therefore, it is natural that our multiple-center integrals are comparable in magnitude with Hubbard's term.

We assume here that the positive ions only play a major role when forming the crystal material at high temperature.   After cooling down to the superconducting temperature, the effect of the positive ions will likely be negligible.   Meanwhile, the p-electron hybrids of the negative ions form the super-molecular cell in the lattice structure.

For illustration, the following discussion will focus on the 2-D $O_4$ lattice in the $CuO_2$ crystals.

II. The Molecular Orbitals of Lattice Crystals

   We revisit the rules of the molecular orbitals (MO) first.
1. The number of MOs formed is equal to the number of atomic orbitals combined.
2. When two atomic orbitals are combined, one of the MOs formed is a bonding MO at a lower energy state than the original atomic orbitals.   The other is an antibonding MO at a higher energy state.
3. In ground-state configurations, electrons enter the lowest energy MOs available.
4. The maximum number of electrons in any given MO is two (Pauli exclusion principle).
5. In ground-state configurations, electrons enter MOs of identical energies singly before they pair up (Hund's rule).

By using the MOs principles to $O_2$, we get the picture of the orbital in Fig. 1,

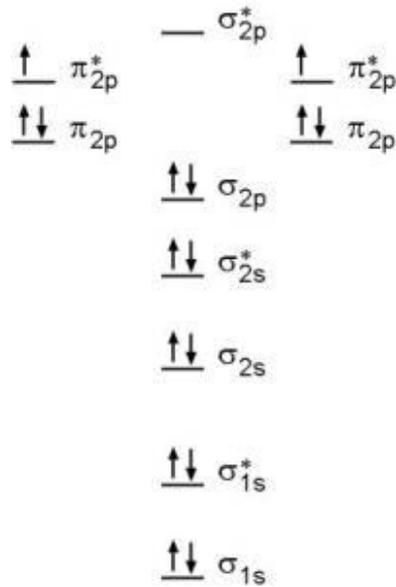

Fig. 1    Orbital diagrams for $O_2$

The orbitals of $O_2$ have valence p-electrons in the two outermost shells.    These electrons are in the antibonding energy level, $\pi^*_{2pz}$ and $\pi^*_{2py}$, as shown in Fig. 1

The Cu 3d energy level is -10.74 ev and the O 2p is -15.9 ev.[2]    The difference between these energy levels is 5.17 ev.    This difference, under basic chemistry principles, would ordinarily be too big for the 2p electrons of O to hybridize with the 3d electrons of Cu at lower temperatures.    However, this would not prevent the 2p O electrons from hybridizing with the other 2p O electrons at this lower temperature.  Initially, we will neglect the presence of the Cu 3d-holes and only consider the presence of the Os in the $CuO_2$ lattice plane.    From the crystal structure of $La_{2-x}Sr_xCuO_4$, each cell of the 2D $CuO_2$ lattice has four O atoms, or $O_4$.    We will find the MOs of one $O_4$ first and then build the MOs of the $O_4$ lattice plane.

Extending the MO rules to the giant molecule in the $O_4$ lattice plane, the MOs of $O_4$ are built from the two $O_2$ MOs, $\pi^*_{2pz}$ and $\pi^*_{2py}$.    The two antibonding MOs, $\pi^*_{2pz}$ and $\pi^*_{2py}$ of one $O_2$ hybridize with the other two MOs $\pi^*_{2pz}$ and $\pi^*_{2py}$ of the other $O_2$. This leads to six different MOs, two from $\pi^*_{2py}$ x $\pi^*_{2py}$, which are lying in the same $CuO_2$ plane, two from $\pi^*_{2pz}$ x $\pi^*_{2pz}$, which are clouds on both sides of the $CuO_2$ plane, and another two from $\pi^*_{2py}$ x $\pi^*_{2pz}$, which lies between the $CuO_2$ plane and the electron cloud.

Following the MO rules, the orbitals with the highest energy states out of these six orbitals are the antibonding orbitals of $\pi^*_{2pz}$ x $\pi^*_{2pz}$ (see Fig. 2).    Since each $O_4$ cell is surrounded by four other $O_4$ cells, the original atomic $p_z$ orbitals of each O atom are perpendicular to the $CuO_2$ plane. The resulting giant molecular antibonding orbital of $\pi^*_{2pz} \times \pi^*_{2pz}$ is a d-like MO.

Fig. 3 shows the 2D-lattice of $CuO_2$ plane, with the $Cu^{+2}$ ions at the points A, B, C, D, and E. The square BCDE is the basic cell with four O atoms in the cell.

From the periodicity, the d-like orbitals of $\pi^*_{2pz} \times \pi^*_{2pz}$ form cloud layers on both sides of the $CuO_2$ plane and have the same pattern and pattern size as the O lattice. The d-like orbitals on each side group up to form two off-plane cloud layers of electrons which provide the physical channels for the superconducting.

The lattice in Fig. 3 can be viewed in a more natural way as the lattice in Fig. 4 since each Cu will be surrounded by four O atoms and each O is shared by two n.n $O_4$ cells. Since each cell is not independent and share the corner O's, valence bond electrons can hop between the neighboring cells. Therefore, for each Cu atom in the $CuO_2$ plane, there are d-like electron clouds of $O_4$ orbital on both sides of the plane. As shown in Fig. 5, the d-like wavefunction is off $CuO_2$ plane, and the orbital pattern and pattern size are same as the O-lattice.

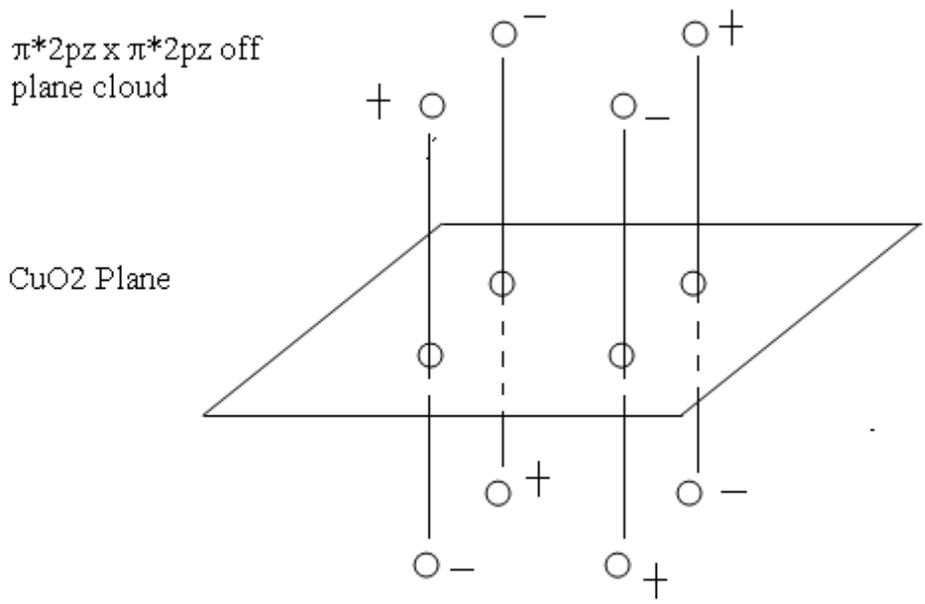

Fig. 2 Four O atoms cell

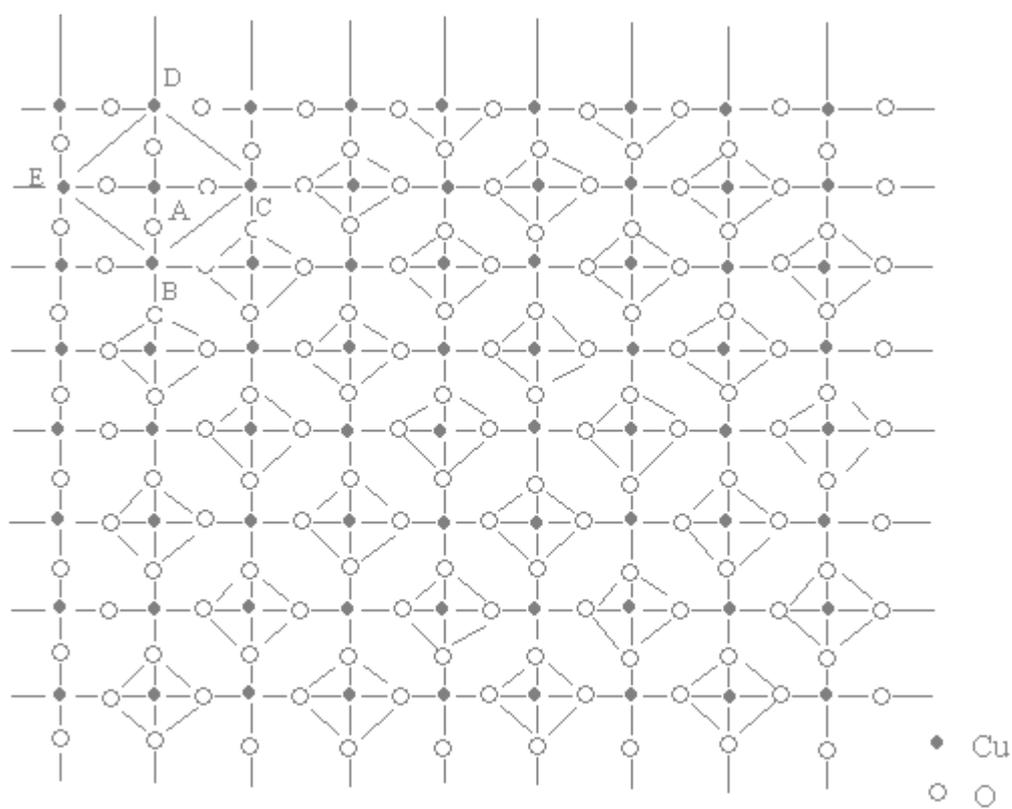

Fig. 3  2D lattice of CuO2

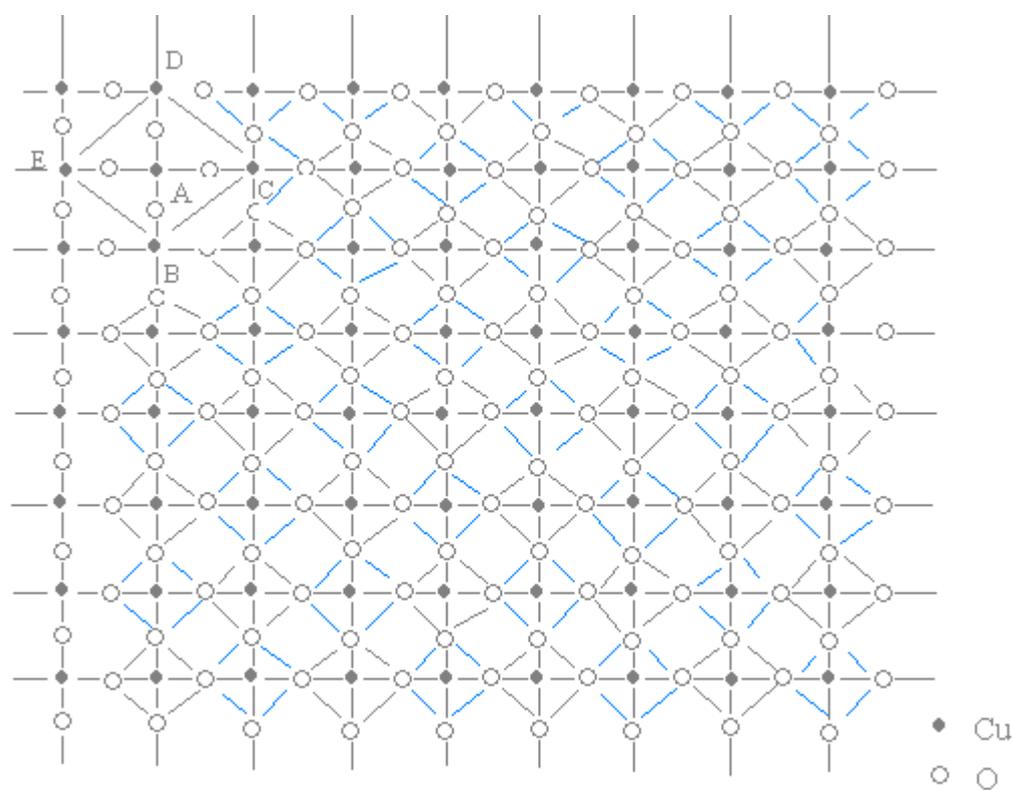

Fig. 4  New view of CuO2 plane

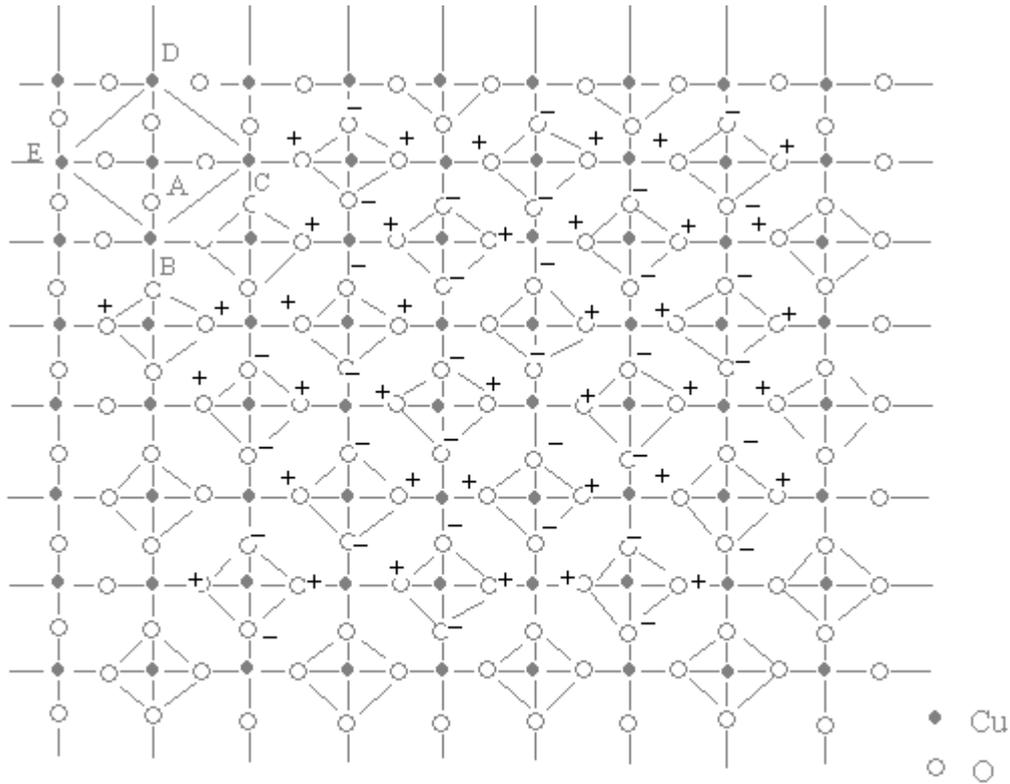

Fig. 5 d-type valence bond electron cloud in 2D periodic O4 cells

+ positive wavefunction
− negative wavefunction

● Cu
○ O

From the periodicity, the d-like wavefunction is as follows;

$\psi(\mathbf{r}) = u(\mathbf{x})\, \phi(z)$

here $\mathbf{r} = (\mathbf{x}, z) = (x,y,z)$

$u(\mathbf{x}) = C \{ \exp(-\alpha((x-L/2)^2 + y^2)) + \exp(-\alpha((x+L/2)^2 + y^2))$
$\qquad - \exp(-\alpha((y-L/2)^2 + x^2)) - \exp(-\alpha((y+L/2)^2 + x^2)) \}$  (1)

here "$\alpha$" is a positive number in the Gauss distribution,

$\phi(z) = z^2/(L/2)^2 \exp\{-z^2/(L/2)^2\}\, [\theta(z) - \theta(-z)]$  (2)

$\Psi(\mathbf{r}) = \Sigma_{\mathbf{R}_l}\, \psi(\mathbf{r} - \mathbf{R}_l)$  (3)

here $\mathbf{R}_l$ 's are the 2D lattice vectors,

$\Psi(\mathbf{r})$ is a 2-D periodic function of $\mathbf{x}$ and anti-symmetric in z direction with respect to the $CuO_2$ plane.

For $La_{2-x}Sr_xCuO_4$, when x=0, no Sr doping, each cell of $O_4$ in the $CuO_2$ plane has six hybridized $\pi^*_{2p} \times \pi^*_{2p}$ orbitals.  To fill all the six orbitals need total twelve electrons. Four electrons come from the two paired $O_2$'s valence bonds, four come from the two $Cu^{+2}$, and another four came from the four $La^{+3}$ 's (total 12 electrons with eight of them being taken by the other four off plane $O^{-2}$ s).  Thus, in $La_2CuO_4$, all the d-like cloud orbitals are filled by pairing electrons as spin singlets with a symmetric space wavefunction, the whole 2-D $O_4$ lattice has a spin vacuum, and the $Cu^{+2}$ lattice is an antiferromagnetic insulator.  The space extension of the $Cu^{+2}$ d-wavefunctions are smaller in size compared to the $O_4$ lattice cell.

When x ≠ 0 and some electrons are missing from the d-like orbitals, there are positive holes in the cloud planes.  Since Hund's Rule prohibits double occupancy of holes in the cells, only one electron may be missing from each cell.  Therefore, the maximum doping is when x = 1/2 when each $O_4$ cell has one hole.  At the optimal doping x = 1/8, one quarter of the cells were single hole occupancy.  Also, when x ≠ 0 , the antiferromagnetic long range order of the $Cu^{+2}$ lattice is destroyed, and the d-holes of the $Cu^{+2}$ will couple with the holes in the O- lattice to form spin singlets locally to maintain the spin vacuum[3,5].

### III. Hubbard Hamiltonian

From the assumption of the valence bond d-like wave function in equation (1), we can derive the Hamiltonian following the method by Hubbard.[4]  The space size of the d-like cloud is the same as that of the $O_4$ cell of the O-lattice.  They are off the $CuO_2$ plane.  Due to the sharing of the corner O atoms by two nearest neighbor cells, the Hubbard Hamiltonian should have more terms than the original Hubbard model.

For a 2N-electrons system of 2D lattice, let h(**r**) be the single electron Hamiltonian in a periodic lattice,

$$H = \Sigma_i\, h(\mathbf{r}_i) + \tfrac{1}{2} \Sigma_{i,j}\, v_{ij} \qquad (4)$$
$$\text{here } v_{ij} = e^2/|\mathbf{r}_i - \mathbf{r}_j|$$

The Bloch function $\psi_\mathbf{k}(\mathbf{r})$, satisfies the following equation;

$$h(\mathbf{r})\,\psi_\mathbf{k}(\mathbf{r}) = E_\mathbf{k}\,\psi_\mathbf{k}(\mathbf{r}) \qquad (5)$$

$H = H_o + H'$

here $H_o = \Sigma_{k,\sigma} E_k C^+_{k\sigma} C_{k\sigma}$   as in the second quantization,

let   $\psi_k(r) = \exp(ikr) \Psi(r)$      with **k** ∈ 1$^{st}$ BZ,

here we choose $\Psi(r)$ as the d-like wavefunction in Equation (3),

$$H' = \tfrac{1}{2} \Sigma_{k1\,k2\,k1'\,k2'} \ \langle k1\, k2 | v | k1'\, k2' \rangle\, C^+_{k1,\sigma} C^+_{k2,\sigma'} C_{k1',\sigma'} C_{k2',\sigma} \quad (6)$$

here
$$\langle k1\, k2 | v | k1'\, k2' \rangle = e^2 \int dr\, dr'\, \psi^*_{k1}(r)\, \psi^*_{k2}(r')\, \psi_{k1'}(r)\, \psi_{k2'}(r') / |r - r'|$$

Then, we introduce the Wannier function $a(r)$,

$$\psi_k(r) = 1/N^{1/2} \Sigma_i \exp(ik R_i)\, a(r - R_i) \quad (7)$$

and   $$a(r - R_i) = 1/N^{1/2} \Sigma_{k \in BZ} \exp(-ik R_i)\, \psi_k(r) \quad (8)$$

$$C^+_{i\sigma} = 1/N^{1/2} \Sigma_{k \in BZ} \exp(-ik R_i)\, C^+_{k,\sigma} \quad (9)$$

The $C^+_{i\sigma}$'s follow the common fermionic anti-commutation relationship.
We can rewrite the H' in Wannier picture as follows;

$$H' = \tfrac{1}{2} \Sigma_{i,j,l,m} \ \langle i\, j | v | l\, m \rangle\, C^+_{i,\sigma} C^+_{j,\sigma'} C_{l,\sigma'} C_{m,\sigma} \quad (10)$$

here
$\langle i\, j | v | l\, m \rangle =$
$$e^2 \int dr\, dr'\, a(r - R_i)^*\, a(r' - R_j)^*\, a(r - R_l)\, a(r' - R_m) / |r - r'| \quad (11)$$

Since here we take $\Psi(r)$ as the d-like wavefunction of equation (1), which is independent of **k** vectors, we can calculate the Wannier function $a(r - R_i)$ as follows;

$$a(r - R_i) = \Psi(r)\, \sin(\pi(x - x_i)/L)\, \sin(\pi(y - y_i)/L) / [\pi(x - x_i)/L)\, (\pi(y - y_i)/L)] \quad (12)$$

The integration of equation (11) has a contribution only when $|r - r'|$ is small, when **r**, **r'** are close to $R_i$ s, and when $u(r)$ is non zero.   Here, we limit the calculation of d**r** d**r**' within the original cell.

Hubbard made an approximation by taking only a single center integral where he let

$R_i = R_j = R_l = R_m = (0,0)$ at the origin point.   The single center integral is

$< i i | v | i i > = e^2 \int d\mathbf{r} \, d\mathbf{r'} \, a(\mathbf{r}- \mathbf{R}_i)^* \, a(\mathbf{r'}- \mathbf{R}_i)^* \, a(\mathbf{r}- \mathbf{R}_i) \, a(\mathbf{r'}- \mathbf{R}_i) \quad /|\mathbf{r}-\mathbf{r'}|$

$= e^2 \int d\mathbf{r} \, d\mathbf{r'} \, \Psi(\mathbf{r})^2 \Psi(\mathbf{r'})^2 \, \{\sin(\pi(x)/L)\sin(\pi(y)/L)/[\pi(x)/L)(\pi(y)/L)]\}^2 \times$

$\{\sin(\pi(x')/L)\sin(\pi(y')/L)/[\pi(x')/L)(\pi(y')/L)]\}^2 \quad /|\mathbf{r}-\mathbf{r'}| \qquad (13)$

$= U > 0$

In our case, there are other non-zero contributions from the multiple-center integration of the nearest neighbors cells as <i,j>,<i, l>, etc.

For example, we can calculate the case when three electrons are in the origin cell and a fourth one in the cell with its center at point (1, 0),

$< 0\, 0 | v | i\, 0> \;= e^2 \int d\mathbf{r} \, d\mathbf{r'} \, \Psi(\mathbf{r})^2 \Psi(\mathbf{r'})^2 \, \{\sin(\pi x/L)\sin(\pi(x-L)/L)\sin^2(\pi y/L)/[(\pi x/L) \times (\pi(x-L)/L)(\pi(y)/L)^2]\} \{\sin(\pi(x')/L)\sin(\pi(y')/L)/[\pi(x')/L)(\pi(y')/L)]\}^2 \quad /|\mathbf{r}-\mathbf{r'}|$

To make a clearer comparison, we assume that the positive parameter $\alpha$ in the d-like wavefunction is very large, as a sharp Gauss distribution, so that the wavefunction has significant contribution only at those small areas around the four points, ( 0, ± ½) and ( ± ½,0 ).   From the integral symmetry, we get,

$< 0\, 0 | v | i\, 0> \;= U/2$

Therefore, the H' should have the following terms;

$U/4 \, \Sigma_{i, <i,j> \sigma} (C^+_{i,\sigma'} C_{j\sigma'} + C^+_{j,\sigma'} C_{i\sigma'}) \, n_{i\sigma}$

where    $n_{i\sigma} = C^+_{i,\sigma} C_{i\sigma}$    and σ' has the opposite spin of σ,

Similar calculations can be made for two-center integrals with two electrons in one cell and the other two in the same n.n cell.   Doing so will give us the following contributions,

$U/3 \, \Sigma_{<i,j> \sigma} \{ C^+_{i,\sigma} C^+_{i,\sigma'} C_{j\sigma'} C_{j\sigma} + C^+_{i,\sigma} C^+_{j,\sigma'} C_{j\sigma'} C_{i\sigma} + C^+_{i,\sigma} C^+_{j,\sigma'} C_{i\sigma'} C_{j\sigma} +$ h.c. $\}$ $\qquad (14)$

We will neglect the other two-center integrals since they make smaller contributions.

For the three-center integrals, we only consider the case when two electrons are in one cell and the other two are in two different n.n cells.    The results are as follows

$$U/8 \sum_{i, <i,j> <i,l> \sigma} \{ (C^+_{l,\sigma'} C_{j\sigma'} + C^+_{j,\sigma'} C_{l\sigma'}) n_{i\sigma} +$$

$$(C^+_{l,\sigma} C_{i\sigma} + C^+_{i,\sigma} C_{l\sigma})(C^+_{i,\sigma'} C_{j\sigma'} + C^+_{j,\sigma'} C_{i\sigma'}) \}$$

Defining $\pi_{ij\sigma} = C^+_{j\sigma} C_{i\sigma} + C^+_{i,\sigma} C_{j\sigma}$

$$H' = U \sum_i n_{i\uparrow} n_{i\downarrow} + U/4 \sum_{i, <i,j> \sigma} \{ \pi_{ij\sigma} \; n_{i\sigma'} \} +$$

$$+ U/3 \sum_{<i,j> \sigma} \{ (\pi_{ij\sigma} \pi_{ij\sigma'}) + n_{i\sigma} n_{j\sigma'} + n_{j\sigma} n_{i\sigma'} \}$$

$$+ U/8 \sum_{i, <i,j> <i,l> \sigma} \{ \pi_{lj\sigma'} \; n_{i\sigma} + \pi_{il\sigma} \pi_{ij\sigma'} \} \qquad (15)$$

Notice here that $\pi^+_{ij\sigma} = \pi_{ij\sigma}$, and $n^+_{i\sigma} = n_{i\sigma}$. They are observable quantum operators.

The first term in H' is the same as the Hubbard Model of single-center integral contribution.

For simplicity, we will first only consider the first few terms in equation (15),

$$H = H_0 + H' = T_0 \sum_i n_{i\sigma} - t_1 \sum_{<i,j> \sigma} \pi_{ij\sigma} + U \sum_i n_{i\uparrow} n_{i\downarrow} + U/4 \sum_{i, <i,j> \sigma} \{ \pi_{ij\sigma} \; n_{i\sigma'} \} +$$

$$+ U/3 \sum_{<i,j> \sigma} \{ (\pi_{ij\sigma} \pi_{ij\sigma'}) + n_{i\sigma} n_{j\sigma'} + n_{j\sigma} n_{i\sigma'} \} \qquad (16)$$

Here $t_1$ is positive.
We can rewrite the equation (16) in more symmetric way,

$$H = T_0 \sum_i n_{i\sigma} - t_1 \sum_{<i,j> \sigma} \pi_{ij\sigma} + U \sum_i n_{i\uparrow} n_{i\downarrow} + U \sum_{<i,j> \sigma} \{ \pi_{ij\sigma} (n_{i\sigma'} + n_{j\sigma'})/2 \} +$$

$$+ U/3 \sum_{<i,j> \sigma} \{ (\pi_{ij\sigma} \pi_{ij\sigma'}) + (n_{i\sigma} n_{j\sigma'} + n_{j\sigma} n_{i\sigma'}) \} \qquad (17)$$

Here $\sum_{<i,j>}$ counts each pair <i,j> only once.

For fixed N cubic lattice sites, there are a total of 2N n.n pairs or edges. If the doping ratio is x, then the number of holes is M= xN. In eq. (17) the first term is a constant since $\Sigma_i n_{i\sigma}$ = M, and can be neglected in the following statistics calculation.

For each site " i ", compare the terms of U $\Sigma_i n_{i\uparrow} n_{i\downarrow}$ and U/3 $\Sigma_{<i,j> \sigma}(n_{i\sigma} n_{j\sigma'} + n_{j\sigma} n_{i\sigma'})$. The ground states will favor lower energy, so the second term will not favor a double occupied state. It will favor one hole with spin up in "I" and another hole with spin down in the n.n site of "j". It is obvious that for each configuration of distribution M holes in N lattice sites, both terms can be calculated by counting the number of pair connections with such spins.

**IV. Phase Transition**

For simplicity, we will consider the three terms in equation (17) as follows,

$$H = -t_1 \Sigma_{<i,j> \sigma} \pi_{ij\sigma} + U \Sigma_{<i,j> \sigma} \{\pi_{ij\sigma}(n_{i\sigma'} + n_{j\sigma'})/2\} +$$

$$+ U/3 \Sigma_{<i,j> \sigma} \{(\pi_{ij\sigma} \pi_{ij\sigma'})\}$$

$$= -t_1 \Sigma_{<i,j> \sigma} \{\pi_{ij\sigma} (1 - (U/2t_1)(n_{i\sigma'} + n_{j\sigma'}))\} +$$

$$+ U/3 \Sigma_{<i,j> \sigma} \{(\pi_{ij\sigma} \pi_{ij\sigma'})\} \qquad (18)$$

We have,
$$(\pi_{ij\sigma})^2 = (C^+_{j\sigma} C_{i\sigma} + C^+_{i,\sigma} C_{j\sigma})^2 = n_{i\sigma} + n_{j\sigma} - 2n_{i\sigma}n_{j\sigma}$$

which will equal zero when both $n_{i\sigma}$ and $n_{j\sigma}$ equal zero or 1, and it will equal 1 when only one of $n_{i\sigma}$ or $n_{j\sigma}$ equals one.

Therefore, for pairs with non-zero terms of $\pi_{ij\sigma}$, $(n_{i\sigma'} + n_{j\sigma'})$ = 1,
So
$$H = -t_1 \Sigma_{<i,j> \sigma} \{\pi_{ij\sigma} (1 - (U/2t_1))\} +$$

$$+ U/3 \Sigma_{<i,j> \sigma} \{(\pi_{ij\sigma} \pi_{ij\sigma'})\} \qquad (19)$$

For the special case if $U = 2t_1$, we will have only one term left as,

$$H = U/3 \Sigma_{<i,j> \sigma} \{(\pi_{ij\sigma} \pi_{ij\sigma'})\} \qquad (20)$$

Let $\Pi_{<ij>\sigma} = \sqrt{(n_{i\sigma} + n_{j\sigma} - 2n_{i\sigma}n_{j\sigma})}$,

For a non-zero $\pi_{ij\sigma}$, $\pi_{ij\sigma} = \Pi_{<ij>\sigma} \exp(i\theta_{<ij>\sigma}) = \pm 1$ as phase $\theta_{<ij>\sigma}$ equals 0 or $\pi$.
For zero $\pi_{ij\sigma}$ we cannot define the phase for the pair $<i,j>$.

We define the pair connection of $<ij>$ open when $\pi_{ij\sigma}$ is non-zero and closed when $\pi_{ij\sigma}$ is zero. According to the edge percolation theory, when the population of the open edge gets to 0.5 for the square 2D lattice, a phase transition for an infinite path of open edges appeared. As mentioned earlier, this will provide the physical path for superconducting electrons.

When temperature T is high, we can expect the $\pi_{ij\sigma}$'s to equal 0 or $\pm 1$, and that the average of H will be zero. But when T becomes smaller, there might be a phase transition of $\theta_{<ij>\sigma} = +1$ for spin up and $\theta_{<ij>\sigma} = -1$ for spin down and H equals the value of $U/3 \Sigma_{<i,j>\sigma} \{(-1)\}$ summed over the site pairs of non-zero $\pi_{ij\sigma}$.

In the general case, equ. (19) can be rewritten as,

$$H = -t_1 \Sigma_{<i,j>\sigma} \{\pi_{ij\sigma}(1 - (U/2t_1))\} + U/3 \Sigma_{<i,j>\sigma} \{(\pi_{ij\sigma} \pi_{ij\sigma'})\}$$

$$= U/3 \Sigma_{<i,j>\sigma} \{(\pi_{ij\sigma} - 3(t_1/U - 1/2))(\pi_{ij\sigma'} - 3(t_1/U - 1/2))\}$$

$$- U/3 \Sigma_{<i,j>\sigma} \{3(t_1/U - 1/2)\}^2$$

The second term is a constant and can be neglected.

$$H = U/3 \Sigma_{<i,j>\sigma} \{(\pi_{ij\sigma} - 3(t_1/U - 1/2))(\pi_{ij\sigma'} - 3(t_1/U - 1/2))\} \quad (21)$$

Since $\pi_{ij\sigma}$ is $\pm 1$, if the absolute value of $|3(t_1/U - 1/2)| > 1$, then H is always positive, and there is no phase transition. Only when $|3(t_1/U - 1/2)| < 1$ might there be a phase transition if the phase becomes ordered for non-zero $\pi_{ij\sigma}$. This will happen only when $1/6 < t_1/U < 5/6$. The phase of the ordered state will be the ground state with a negative energy. This is the phase transition of superconductivity.

The brackets designate the canonical average

$$<A> = Tr(\rho A)$$

$$\rho = e^{-\beta H} / Tr(e^{-\beta H})$$

where β is the inverse of the temperature and the trace is performed over the Hilbert space with a fixed N number of electrons.

If we use equ. (20) for H to calculate the average value of $\pi_{ij\sigma}$,

$$<\pi_{ij\sigma}> = Tr(\rho \pi_{ij\sigma})$$

When N and M are large enough or tend towards infinity, then each $\pi_{ij\sigma}$ will be independently averaged.

$$<\pi_{ij\sigma}> = (\Sigma_{config}\, \pi_{ij\sigma} \exp(-\beta E^{config})) / (\Sigma_{config}\, \exp(-\beta E^{config}))$$
$$= ((+1)\exp(-\beta \pi_{ij\sigma'}) + (-1)\exp(\beta \pi_{ij\sigma'})) / (2 + \exp(-\beta \pi_{ij\sigma'}) + \exp(\beta \pi_{ij\sigma'})) \quad (22)$$

Here, the factor 2 comes in because there are two cases of zero value for $\pi_{ij\sigma}$ when the sites of i and j are both nulls or occupied at the same time.  The above equation is correct only when the following conditions apply;

$$\exp(-\beta E^{config}) = \exp(-\beta U/3 \Sigma_{<i,j>\,\sigma} \{(\pi_{ij\sigma}^{config} \pi_{ij\sigma'}^{config})\})$$

$$= \exp(-\beta U/3\, (\pi_{ij\sigma}^{config} \pi_{ij\sigma'}^{config}))\exp(-\beta U/3 \Sigma'_{<i,j>\,'\sigma}\{(\pi_{ij'\sigma}^{config} \pi_{ij'\sigma'}^{config})\}) \quad (23)$$

and
$$\exp(-\beta U/3 \Sigma'_{<i,j>'\,\sigma}\{(\pi_{ij'\sigma}^{config} \pi_{ij'\sigma'}^{config})\}) \text{ (at } \pi_{ij\sigma}^{config} = 1 \text{ or } 0\text{)}$$

$$= \exp(-\beta U/3 \Sigma'_{<i,j>'\,\sigma}\{(\pi_{ij'\sigma}^{config} \pi_{ij'\sigma'}^{config})\}) \text{ (at } \pi_{ij\sigma}^{config} = -1 \text{ or } 0\text{)}$$

The summation $\Sigma'_{<i,j>'\,\sigma}$ in the second factor of exp ( ) is over all the pairs except the pair of <i,j>.  When the N and M are large enough will we expect that the conditions (23) applied.

From equ. (22) we have,

$$<\pi_{ij\sigma}> = (\exp(-\beta \pi_{ij\sigma'}/2) - \exp(\beta \pi_{ij\sigma'}/2)) / (\exp(-\beta \pi_{ij\sigma'}/2) + \exp(\beta \pi_{ij\sigma'}/2))$$

$$= (1 - \exp(\beta \pi_{ij\sigma'})) / (1 + \exp(\beta \pi_{ij\sigma'})) \quad (24)$$

From (24), if $\pi_{ij\sigma'} = 0$, then $\langle\pi_{ij\sigma}\rangle = 0$ at any finite temperature.
But when T tends to zero, if $\pi_{ij\sigma'} = 1$, $\langle\pi_{ij\sigma}\rangle = -1$ and if $\pi_{ij\sigma'} = -1$, $\langle\pi_{ij\sigma}\rangle = 1$

This also confirms that at a low enough temperature, $\pi_{ij\sigma}$ is favored to have the opposite signs of phases. This is the mechanism for the ordered phase transition. As a result of the periodicity, we expect this result to be true for each pair of <i,j>. This critical temperature $T_c$ can be found by using a computer simulation for all the different configurations when N and M are given.

From symmetry, we can expect the number of pair connection with spin up will equal with those of spin down. Randomly distributed M holes to N sites will be easily clogged if M/N is too large (about 0.5), and would not be able to form an infinity path if M/N is too small. For both situations there will be no superconducting transition. From the percolation, it is quite natural to explain the dome of $T_c$ with the doping and the sign change of the Hall currents.

When applyed to the 3D case, we notice that each itinerant electron hopping will have a phase change of 0 or π. Therefore, we can define the winding number for the closed path in the 3D as integer factor multiple of 2π for the total phase change of the connected chain.

We can also calculate that,

$$Tr \{ (\exp(-\beta U/3 \Sigma_{<i,j>\sigma} (\pi_{ij\sigma}\pi_{ij\sigma'}))) C^+_{L,\uparrow} C^+_{L,\downarrow} C_{M,\downarrow} C_{M,\uparrow} \} \text{ is non zero}$$

while $|L - M| \longrightarrow$ infinite,

which will give $\langle C^+_{i,\uparrow} C^+_{i,\downarrow} C_{j,\downarrow} C_{j,\uparrow} \rangle \neq 0$ when $|i - j| \to \infty$

This is the Off-Diagonal Long-Range Orders[6].

From this model, we believe that a Graphene monolayer will also show superconductivity in low temperatures if the suitable doping of holes can be realized. Graphene has the same hexagonal lattice structure. $KC_8$ is a superconductor and $LiC_6$ is not[7] because the former has a doping of 0.11 "grapheme $\pi^*$ electron" per grapheme unit cell while the later has only 0.0344 electrons. From the percolation theory, it is obvious that the pair connection in LiC$_6$ is not enough to have the infinite network available for a superconducting path.